\documentclass[useAMS,usenatbib]{mn2e}
\usepackage{epsfig}
\usepackage{deluxetable}
\usepackage{graphicx}
\usepackage{natbib}

\newcommand\ion[2]{#1{\scshape{#2}}}%

\newcommand{\lsol}{L$_{\odot}$}
\newcommand{\lir}{$L_{\rm IR}$}
\newcommand{\ms}{$M_{\ast}$}
\newcommand{\lco}{$L^{\prime}_{\rm CO}$}

\newcommand{\lcii}{$L_{\rm CII}$}
\newcommand{\Mmol}{$M_{\rm H_{\rm 2}}$}
\newcommand{\ha}{H$\rm \alpha$}
\newcommand{\re}{$r_{\rm 1/2}$(H$\alpha$)}
\newcommand{\rei}{$r^{\prime}_{\rm 1/2}$}
\newcommand{\den}{$\rm \Sigma_{SFR}$}
\newcommand{\mden}{$\rm \Sigma_{\ast}$}

\title[Mapping the \ha\ emission of star forming galaxies at $z\sim1$]{KROSS: Mapping the \ha\ emission across the star-formation sequence at $z \approx 1$}
\author[Georgios E. Magdis et al.]
  {Georgios E. Magdis$^{1,2,3}$\thanks{E-mail: magdis@dark-cosmology.dk}, Martin Bureau$^{1}$, J. P. Stott$^{1,4}$, A. Tiley$^1$, A. M. Swinbank$^{5,4}$,\newauthor R. Bower$^{4,5}$, A. J. Bunker$^{1,6}$, Matt Jarvis$^{1,7}$, Helen Johnson$^{5}$, Ray Sharples$^{4,8}$
\\  \\
  $^1$Department of Physics, University of Oxford, Keble Road, Oxford OX1 3RH, UK\\
  $^2$Dark Cosmology Centre, Niels Bohr Institute, University of Copenhagen, Juliane Mariesvej 30, DK-2100 Copenhagen\\
  $^3$Institute for Astronomy, Astrophysics, Space Applications and Remote Sensing, National Observatory of Athens, GR-15236 Athens, Greece\\
  $^4$Institute for Computational Cosmology, Durham University, South Road, Durham DH1 3LE, UK\\
  $^5$Centre for Extragalactic Astronomy, Department of Physics, Durham University, South Road, Durham DH1 3LE, UK\\
  $^6$Affiliate Member, Kavli Institute for the Physics and Mathematics of the Universe, 5-1-5 Kashiwanoha, Kashiwa, 277-8583, Japan\\
  $^7$Department of Physics, University of the Western Cape, Bellville 7535, South Africa\\
  $^8$Centre for Advanced Instrumentation, Department of Physics, Durham University, South Road, Durham, DH1 3LE, UK}
\begin{document}

\date{}

\pagerange{\pageref{firstpage}--\pageref{lastpage}} \pubyear{2002}

\maketitle

\label{firstpage}

\begin{abstract}
We present first results from the KMOS Redshift One Spectroscopic Survey (KROSS),  an ongoing large kinematical survey of a thousand, $z \sim 1$ star forming galaxies, with VLT KMOS. Out of the targeted galaxies ($\sim$ 500 so far), we detect and spatially resolve \ha\ emission in $\sim 90\%$ and 77\% of the sample respectively. Based on the integrated \ha\ flux measurements and the spatially resolved maps we derive a median star formation rate (SFR) of $\sim$ 7.0\,M$_{\odot}$\,yr$^{-1}$ and a median physical size  of $\langle$\rei$\rangle$ $=$ 5.1\,kpc. We combine the inferred SFRs and effective radii measurements to derive the star formation surface densities (\den) and present a ``resolved" version of the star formation main sequence (MS) that appears 
to hold at sub-galactic scales, with similar slope and scatter as the one inferred from galaxy integrated properties.   Our data also yield a trend between \den\ and  $\rm \Delta(sSFR)$ (distance from the MS) suggesting that galaxies with higher sSFR are characterised by denser star formation activity.  Similarly,  we find evidence for an anti-correlation between the gas phase metallicity (Z) and the $\rm \Delta(sSFR)$, suggesting a 0.2\,dex variation in the metal content of galaxies within the MS and significantly lower metallicities for galaxies above it. The origin of the observed trends between \den $-$ $\rm \Delta(sSFR)$ and Z $-$ $\rm \Delta(sSFR)$ could be driven by an interplay between variations of the gas fraction or the star formation efficiency of the galaxies along and off the MS. To address this, follow-up observations of the our sample that will allow gas mass estimates are necessary
  
\end{abstract}


\section{Introduction}
A recent major step forward in understanding the nature of star formation in distant galaxies has been the discovery that the majority of star-forming galaxies show a strong correlation between their star formation rate (SFR) and their stellar mass ($M_{\ast}$)  from $z = 0$ to $z = 7$ (e.g. Brinchmann et al.\,2004; Noeske et al.\,2007; Elbaz et al.\,2007; Daddi et al.\,2007; Pannella et al.\,2009; Magdis et al.\,2010; Whitaker et al.\,2014). This correlation,  the star-formation sequence or the main sequence (MS) of star formation, is characterised by an increasing normalisation factor (parametrised by the specific star formation rate, sSFR $\equiv$ SFR/$M_{\ast}$) with look back time (e.g. Gonzalez et al.\,2010; Elbaz et al.\,2011) and  a rather small and constant scatter of 0.3\,dex  at all redshifts and stellar masses (e.g. Schreiber et al.\,2015).  The tightness of this correlation, that speaks against stochastic merger-induced star-forming episodes as the main driver of star formation activity, has been put forward as indirect evidence for secular galaxy evolution (e.g. Noeske et al.\,2007).

The contrasting nature of MS galaxies to that of galaxies that, due to their enhanced sSFRs, lie above the MS (hereafter starburst galaxies), is manifested in terms of  various sets of observables. As galaxies depart 
from the MS, they appear to have higher total-to-mid-IR luminosity ratio, IR8 $\equiv$ \lir/$L_{\rm8}$ (Elbaz et al.\,2011; Nordon et al.\,2012, where \lir\ and $L_{\rm 8}$ are the total infrared and the rest frame 8\,$\mu$m luminosity, respectively), weaker far-IR atomic lines (e.g. Garcia-Carpio et al.\,2011; De Looze et al.\,2014; Magdis et al.\,2014), higher star formation efficiencies (SFE $\equiv$ SFR/\Mmol, Daddi et al.\,2010; Genzel et al.\,2010; Magdis et al.\,2012b), and higher dust temperatures (e.g. Magdis et al.\,2012b; Magnelli et al.\,2014). Furthermore, morphological studies have shown a trend of increasing incidence of interacting/merging systems as a function of distance from the MS (Stott et al.\,2013, Hung et al.\,2013). In that respect, there is a trend towards a new paradigm where the properties of the galaxies are not, as traditionally thought, just linked to their IR luminosities (or SFRs), but rather to their specific star formation rate ($ \rm sSFR=SFR/M_{\ast}$), or equally, to their position with respect to the main sequence.

The observed deviations of starbursts from the SFR$-$ $M_{\ast}$, \lir\ $-$ $L_{\rm 8}$, \lir\ $-$ \lcii\ and \lir\ $-$ \lco\ scaling laws, that are followed by the majority of star forming galaxies at all redshifts and luminosities, have been interpreted as evidence of more intense star formation, more compact geometries and higher star formation surface densities, probably triggered by  merger events (e.g., Elbaz et al.\,2011, Noeske et al.\,2007, Diaz-Santos et al.\,2013, Magdis et al.\,2014, Daddi et al.\,2010, Genzel et al.\,2010). In particular, Elbaz et al.\,(2011), using radio and mid-IR observations,  found a clear correlation between  the projected star formation rate surface density ($\Sigma_{\rm SFR}$) of local galaxies  and  their distance from the MS. Similarly, using stacking techniques, they showed that $z \approx 0.8-1.5$ starbursts are characterised by high, on average, IR surface brightness 
($\Sigma_{\rm IR} > 3.0 \times 10^{10}$ \lsol\ kpc$^{-2}$), suggestive of compact star formation activity (Diaz-Santos et al.\,2010). 

However,  how (and if) the  $\Sigma_{\rm SFR}$, as well as other physical properties such as the metal abundance of distant galaxies, varies within and outside the main sequence is still unclear.  Here, we attempt to address this question, taking advantage of spatially resolved \ha\ observations of a large sample of z $\approx$ 1.0 star-forming galaxies, as part of the ongoing KMOS Redshift One Spectroscopic Survey (KROSS). Throughout the paper we adopt the WMAP-7 cosmology ($\Omega_{\rm m} = 0.273$, $\Omega_{\rm \Lambda} +  \Omega_{\rm m} = 1.0$,  $\Omega_{\rm k} = 0$ and $H_{\rm 0} =$ 70.4 km\ s$^{-1}$\ Mpc$^{-1}$ ; Larson et al. 2011) and a Chabrier\,(2003) IMF.

\section{Sample, Observations and Data reduction}
KROSS is an ongoing study of $\sim$ 1000 mass-selected star-forming galaxies. The majority of the galaxies in the sample are selected to be brighter than a magnitude limit of $K_{\rm AB}$  $\approx$ 22.5 (that roughly corresponds to a limiting stellar mass of  log($M_{\ast}/M_{\odot}$) $>$ 9.3 $\pm$ 0.5), excluding sources identified as AGNs (either spectroscopically or from their X-ray emission). Other than that, no other colour, or size cut, criterion was applied to the sample. While the  redshift range of the sample spans from  $z \approx 0.7$  to $z \approx 1.5$, 96\% of the galaxies lie within $0.8 < z < 1.0$, with a median of 0.86. The KROSS galaxies are selected from spectroscopic surveys in some of the most well studied fields in thy sky. In particular, targets were drawn from the UDS (Ultra Deep Survey; Lawrence et al.\, 2007), ECDFS (Extended Chandra Deep Field South, Lehmer et al.\,2005), COSMOS (Cosmological Evolution Survey, Scoville et al.\,2007) and SA22 (Steidel et al.\,1998). The number of sources from each field, as well as a reference to the corresponding spectroscopic survey, are listed in Table 1.  We note that the selected galaxies consist a representative sample of typical star forming galaxies at $ z\sim 1 $,  exhibiting a similar distribution in colour (r-z), stellar mass (\ms) and SFR  to that of the parent sample (see Stott et al.\,2015). 

The KROSS survey commenced in November 2013, exploiting the recently commissioned K-band Multi-Object Spectrograph (KMOS, Sharples et al.\,2013), a near-infrared integral field spectrograph on the VLT. The main aim of the survey is to target, primarily, the redshifted \ha\ and [NII] emission lines of $z \approx 1$ star-forming galaxies. A detailed description of the sample selection, the survey design and the data reduction is presented in Stott et al.\,(2015) and Tiley et al.\,(2015, submitted),  

\begin{table*}
\begin{center}
\caption[]{A list of the extragalactic fields observed by KROSS and the parent spectroscopic redshift catalogues from which we source our KMOS targets.}
\label{tab:fields}
\small\begin{tabular}{lllll}

\hline
Field: & UDS&ECDFS&COSMOS&SA22\\
Coordinates (J2000): & 02:17:48 -05:05:45 & 03:32:28 -27:48:30&10:00:28 +02:12:21&22:17:00 00:20:00\\
\hline
Redshift surveys\\
&UDS $^1$&MUSYC$^2$&zCOSMOS$^3$&VVDS$^4$\\
&VIPERS$^5$&VVDS$^4$&HiZELS$^6$&VIPERS$^5$\\

&HiZELS$^6$&...&...&CF-HiZELS$^7$\\
\hline
Nmaster &7168 & 318 &1743 & 8283 \\
Nobs      & 149  & 120  &102  &  116\\
\hline
\multicolumn{5}{l}{$^1$ Smail et al.\,(2008), Akiyama et al. (in prep) and Simpson et al. (in prep)}\\
\multicolumn{5}{l}{$^2$ Cardamone et al.\,(2010) and references therein}\\
\multicolumn{5}{l}{$^3$ Lilly et al.\,(2007)}\\
\multicolumn{5}{l}{$^4$ Le F{\`e}vre et al.\,(2005, 2013) and Garilli et al.\,(2008)}\\
\multicolumn{5}{l}{$^5$ Garilli et al.\,(2014) and Guzzo et al.\,(2014)}\\
\multicolumn{5}{l}{$^6$ Sobral et al.\,(2012, 2013)}\\
\multicolumn{5}{l}{$^7$ Sobral et al.\,(2015, 2013) and Stott et al.\,(2014)}\\
\end{tabular}

\end{center}
\end{table*}

In this work, we employ data from the first 21 fields observed by KMOS as part of the KROSS survey, from November 2013 to November 2014. During this period  KMOS observed a total of 487 galaxies, employing 24 integral field units (IFUs) per field, each with a square field of view of 2.8'' $\times$ 2.8''. The light from each set of 8 IFUs, that are mounted in configurable arms with a patrol field of 7'.2 in diameter, is dispersed in three cryogenic grating spectrometers. Our observations were carried out using the YJ filter with a spectral coverage of 1.01 $-$ 1.35$\,\mu$m at a spectral resolution of $\lambda/\Delta\lambda$ = 3300. We adopted a  ABA (object-sky-object) observing sequence, with 600s integration per pointing. The total on-source integration time was typically 2.5\,hr per galaxy. 
\begin{figure}
\centering
\includegraphics[scale=0.39]{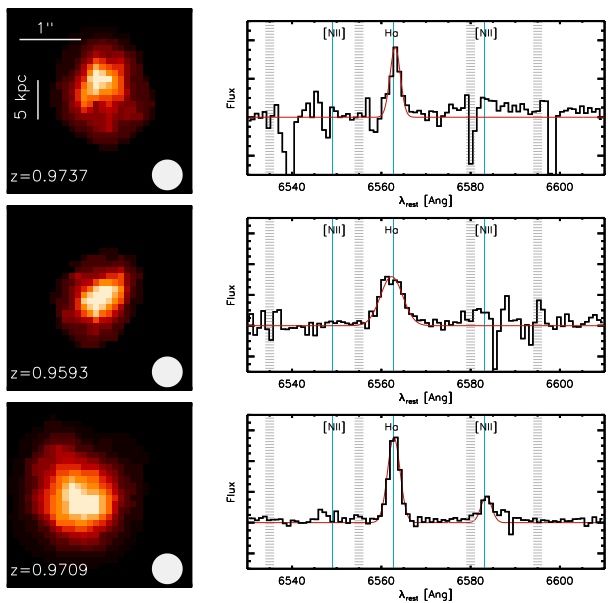}\\
\caption{\ha\ emission map and THE corresponding spectrum for three sources from our sample (below, on and above the MS). The spectra are truncated,  showing only the the wavelength range that covers the [NII]\,6548.03\,\AA, \ha\ and [NII]\,6583\,\AA\ lines, denoted with cyan vertical lines. The grey lines correspond to the locations of bright night sky lines, with a width that corresponds to the FWHM of the effective spectral resolution of the data. The thin red lines depict the best fit line profiles.The x-axis in the spectra has been shifted to the rest-frame based on the measured redshift of each source. The grey circles on the maps, depict the PSF FWHM.}
\label{fig:map} %
\end{figure}

For data reduction we employed the ESOREX/SPARK (Davies et al.\,2013). The software extracts the slices for each IFU and preforms flat$-$fielding, illumination correction and wavelength calibration of the data. Sky OH subtraction was optimised  using a two-step approach. First we subtracted from each object IFU (A), its adjacent sky frame (B). Then, in order to improve the sky subtraction, for each pair of object minus sky (A-B) frames, we further subtracted the average residual between two sky positions. This technique was found to minimise the sky residuals from the final science products. For a detailed description of this method as well as alternative techniques to optimise the sky subtraction for KMOS data we refer the reader to Stott et al.\,(2015). The final (continuum subtracted) \ha\ map for each galaxy was subsequently created by centroiding wavelength-collapsed images around the redshifted \ha\ emission and combining each exposure (i.e. each data cube) using a clipped average.  Both the effects of instrumental resolution and the spatial point speed function (PSF) were taken into account throughout the analysis and are included in the error estimation. Examples of spatially-resolved \ha\ maps and the corresponding rest-frame integrated spectra, centred at redshifted \ha\ emission line, are shown in Figure \ref{fig:map}.

The data of the acquisition and telluric standard stars were reduced following the same procedure as for the science data. During each observation, typically one IFU was placed on a star to monitor the shape and variation of the PSF during the exposure. For each target, the spatial PSF was measured on the two-dimensional images of the acquisition and science stars, obtained by averaging (with $\sigma$-clipping) all the wavelength channels of the stars' reduced data cubes, by fitting  Gaussian profiles. The effective angular resolution of our data sets from the PSF half width at half maximum  (HWHM) ranges from 0.23'' to 0.42'', with a mean of 0.31'' (corresponding to $\sim$2$-$3.4\,kpc and $\sim$2.5\,kpc, respectively, at $z \approx$ 1). 

We fit the \ha\ 6563\AA\ and [\ion{N}{ii}]\,6583\AA\ emission lines allowing  the centroid, intensity and width of a Gaussian profile to vary as a 
free parameter (while keeping the central wavelength and the FWHM of the \ha\ and [\ion{N}{ii}]\,6583\AA\ lines coupled in the fit). With this 
method we derived central wavelengths and integrated fluxes, while the corresponding uncertainties were estimated by perturbing the data 
points within their error bars, and  repeating the fit. Overall we detected integrated \ha\ 6563\AA\ emission (S/N$>$3) in 437 galaxies, 
of which 374  are also spatially resolved, while  [\ion{N}{ii}]\,6583\AA\ emission was detected in 190 galaxies. These numbers yield an \ha\ 
detection and resolving efficiency of $\approx$\,90\% and $\approx$\,77\% respectively, for the KROSS survey. In $\sim$4\% of the sample we 
only detect continuum emission while $\sim$ 6\% remain undetected. The median $r-z$ colour of the galaxies with resolved \ha\ emission is 
0.90 while for those where \ha\ emission remains unresolved  is 1.27. Furthermore, the median $K-$band magnitude for these two subsamples is 21.2 and 20.6 respectively. Finally, galaxies with no \ha\ detection or galaxies where \ha\ emission is unresolved have significantly lower star 
formation rates (as derived from SED fitting, see subsection 3.1) with a median value of $\sim$1.5\,$\rm M_{\odot}$\,$\rm yr^{-1}$, which is 4$-$5 times lower in comparison to the median SFR of galaxies with \ha\ detection ($\sim$7.0\,$\rm M_{\odot}$\,$\rm yr^{-1}$).
The redder $r-z$ colour, the brighter $K-$band magnitude and the significantly lower SFRs  of the \ha\ unresolved/undetected population demonstrates that its the more massive and more passive galaxies that are characterised by weak \ha\ emission (see Table 2). 
\begin{table}
\begin{center}
\caption[]{Statistics of the properties for various subsamples of the observed KROSS galaxies.}
\label{tab:fields}
\small\begin{tabular}{lcccc}
&N& $\rm <(r-z)>$& $\rm <K>$& $\rm <SFR_{SED}>$\\
\hline
Total& 487& 0.98& 21.1& 4.8\\
\hline
\ha    & 437& 0.95& 21.2& 6.8\\
\ha\ resolved & 374& 0.90& 21.2& 7.2\\
\ha\ unresolved & 63& 1.27& 20.6& 2.0\\
\ha\ undetected& 50& 1.32& 20.5& 1.2\\
\hline
[\ion{N}{ii}] (6583\,\AA) & 190& 0.95& 21.3& 6.0\\
\hline
\end{tabular}
\end{center}
\end{table}
\begin{figure*}
\centering
\includegraphics[scale=0.55]{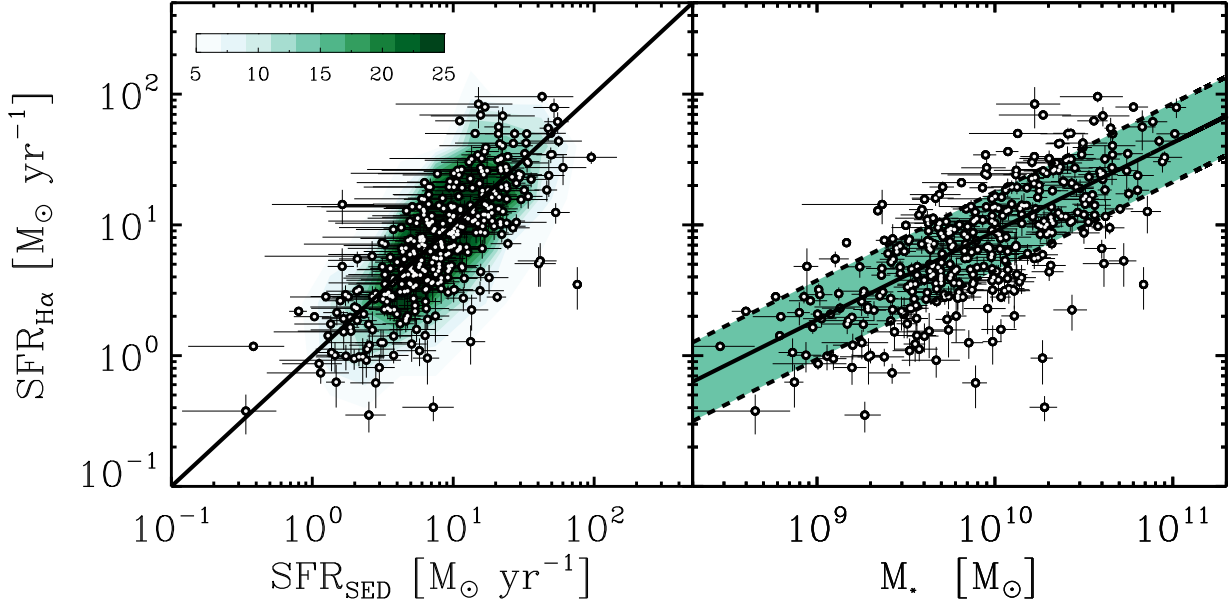}\\
\caption{\textbf{Left)} Comparison of star formation rates as derived based on SED fitting and based on \ha\ flux measurements corrected for dust attenuation. The green contours depict the number density of the sources while the  solid black line corresponds to the 1$-$1 correlation. \textbf{Right)} Star formation rates (based on dust corrected \ha\ measurements) versus the stellar masses of the galaxies in our sample. The solid line represents the star formation sequence at the median redshift of the sample ($z=0.86$) along with a 0.3 scatter (dashed lines) as derived by Speagle et al.\,(2014), taking into account both the dependence  on  redshift and stellar mass. Evidently, our sources follow closely the SFR-M$_{\ast}$ MS at this redshift.}
\label{fig:sfr} %
\end{figure*}
\section{Analysis}
\subsection{Stellar masses, attenuation and star formation rates}
Our targets benefit from extensive rest frame UV to mid-IR photometric coverage. We built a multi-wavelength catalog ($u,B,V,R,I,z,J,K$ + IRAC) using a 2".0 and 3".8 diameter aperture matched photometry in the optical/near-IR and IRAC bands respectively (Cirasuolo et al.\,2007; Williams et al.\,2009; Cardamone et al.\,2010; Kim et al.\,2011b; Muzzin et al.\,2013b; Simpson et al.\,2014).  The photometric bands were consistent between the fields except for SA22 where $g$ was used instead of $B$ and $V$ and no suitable IRAC imaging was available. 

The optical to near-IR data were then used to perform spectra energy distribution (SED) fitting to estimate each galaxy's stellar mass ($M_{\ast}$), dust extinction (A$_{\rm V}$) and age. For this task, we employed the Cigale code (Burgarella et al.\,2005), which fits photometric data points with synthetic stellar population models, and picks the best-fit parameters by minimising the $\chi^{2}$ between observed and model fluxes. The models were built based on the Bruzual and Charlot\,(2003) library assuming a solar metallicity, a constant star formation history, a Calzetti\,(2000) attenuation law and a  minimum age of 0.3\,Gyrs. The stellar mass is obtained by integrating the modelled SFR over the galaxy age, and correcting for the mass loss in the course of stellar evolution. As the derived parameters, especially SFR, are known to be sensitive to the choice of the star formation history (e.g. Mancini et al.\,2011), we repeated the analysis assuming a double exponentially rising SFH. While the absolute values of the derived parameters do vary between the two SFHs, the overall trends reported in this work remain unaffected. In the rest of the paper we use the stellar masses and A$_{\rm V}$ values derived assuming a constant SFH. 

Repeating the fit including this time the AGN module of Cigale in order to quantify the fraction of emission arising from possible AGN activity, does not change the derived properties. This was somewhat expected as from the  parent sample we specifically excluded sources identified as AGNs, either spectroscopically or from their X-ray emission. Also, based on the 
[\ion{N}{ii}]/H$\alpha$ flux ratio AGN diagnostics  of Kewley et al.\,(2001), we find that none of the KROSS galaxies are 
identified as AGN as all have log([\ion{N}{ii}]/H$\alpha$) $<$ 0.2 (of which only 9 have 0.0 $<$ log([\ion{N}{ii}]/H$\alpha$) $<$ 0.2). 

Rather than using the indicative SFR derived through SED fitting, we measured the instantaneous SFR directly from the data using the \ha\ line measurements. The first step was to infer the  observed \ha\ luminosity, $L_{\rm H{\alpha},obs}$, of each galaxy using the \ha\ flux extracted from the spatially integrated spectra. Since most of \ha\ emission is likely to originate from the  H\,II regions that suffer from dust extinction, we derived intrinsic \ha\ luminosity, $L_{\rm H{\alpha},int}$, taking into account the stellar reddening A$_{\rm V}$. In particular, following Wuyts et al.\,(2013), we converted the inferred stellar extinction to gas extinction ($A_{\rm gas}$) and derived the corresponding $L$(\ha, int) through: 
 
\begin{equation}
\begin{centering}
{A_{gas} = A_{V} \times (1.9 - 0.15 A_{ V})},  
\end{centering}
\end{equation} 

\begin{equation}
\begin{centering}
{L_{\rm H\alpha,obs} = L_{\rm H\alpha,int} \times 10^{-0.4A_{\rm gas}}}.
\end{centering}
\end{equation} 
Subsequently, we converted the intrinsic \ha\ luminosities to SFRs using the Kennicutt\,(1998) relation, scaled to a Chabrier IMF (Chabrier 2003):
\begin{equation}
\begin{centering}
{  {\rm log(SFR) [M_{\odot} yr^{-1}]} = {\rm log}(L_{\rm H\alpha,int})\rm{[erg\ s^{-1}]} - 41.33}.
\end{centering}
\end{equation} 
The derived SFRs range from 0.15 to 330\,M$_{\odot}$\,yr$^{-1}$ with a median of $\sim$ 7.0\,M$_{\odot}$\,yr$^{-1}$. A comparison between the \ha\ based SFRs and those obtained through the SED fitting is shown in Figure \ref{fig:sfr}(left), revealing an  overall agreement between the two independent estimates.  

In Figure \ref{fig:sfr}(right), we also place our sources in the SFR-$M_{\ast}$ plane along with the MS at the median redshift of our sample ($z=0.86$) as prescribed by Speagle et al.\,(2014), taking into account the dependence of the MS on both the redshift and stellar mass of the galaxies. Evidently, the majority of our sources exhibit sSFR consistent with that of MS galaxies (within a factor of $\approx$ 2), with a few outliers both below and above the MS. This further demonstrates that the KROSS sample is representative of the population of typical star forming galaxies at $z \sim 1$.

Finally,  we tested a possible bias in our analysis introduced by the more modest sampling of the SED in SA22. First, we repeated whole analysis presented in this work but excluding this time the sub-sample of galaxies  in SA22. Then we repeated the analysis only for galaxies that lie in SA22. In both cases, the trends and general conclusions remain unaffected with respect to those derived when using the whole sample.

\subsection{Size estimates}
To estimate the intrinsic \ha\ sizes of the sources (\rei), we first determined the half-light radius of the \ha\ maps (\re) using  the curve-of-growth analysis in elliptical apertures centred at the dynamical centre of the galaxies, as obtained by modeling the two-dimensional velocity field of each galaxy using a six parameter model (inclination, disk rotation speed, position angle, [x / y] center and rotation curve shape, see Swinbank et al.\,2012 for details). To derive the intrinsic sizes, \rei, we then corrected the measured \re\ for the effective spatial resolution by subtracting in quadrature the PSF half-width at half-maximum (HWHM) appropriate for each data set. As the shape of the PSF is known to vary among the three KMOS spectrographs we derived an average PSF for each spectrograph using the acquisition and the monitor stars at the appropriate IFUs. 

For the measurement of intrinsic size of the galaxies we identify three sources of uncertainty. The first is the uncertainty associated with the size measurement of the galaxies through the growth of curve fitting. The second is the PSF variation during the observations of a particular field. Inspection of the PSFs associated with individual OBs of a given object suggests typical seeing variations of $<$20\%. Finally, the third source of uncertainty arises from the fact that in our analysis we assume that the PSF is axisymmetric, which is not always the case. Indeed, fitting a two-dimensional elliptical gaussian to the PSFs we find that the ellipticity of the PSFs varied between 0.05 and 0.1 for spectrographs one and two and betwee  0.1 and 0.2 for spectrograph three. To derive the uncertainties of the intrinsic (physical) sizes we combine in quadrature the aforementioned errors, that are associated with the growth of curve analysis and the shape of the PSF. The average intrinsic size on our sample is 5.4\,kpc (ranging from 1.4 to 22.3\,kpc) with a corresponding  uncertainty of $\approx$ 30\%. An obvious caveat of our analysis is that due to the seeing limited nature of our observations, we cannot spatially resolve individual star forming clumps in our galaxies. Therefore, we cannot disentangle between a smooth disk-like distribution of the star-formation activity and individual clumpy [HII] regions that are separated by a distance comparable to the size of our PSF.

To account for \ha\ flux missed due to low S/N at the outskirts of each galaxy we also modelled the curve of growth of each galaxy with a Gaussian radial distribution. In almost all cases $>$90\% of the extrapolated, integrated \ha\ flux is recovered by our data.  As a sanity check, we also repeated the size measurements using two independent methods: first by using as centre of the curve of growth the centroid of the line emission as determined from the line maps, and second employing GALFIT to model the radial distribution of the \ha\ flux. The size measurements of the three methods are  consistent within the uncertainties, without any obvious systematics. For the rest of this work, we thus adopt the size measurements based on the curve of growth analysis using the dynamical centre of the galaxies.

The analysis described above was applied to the whole sample of galaxies with \ha\ detection (437). A galaxy was considered resolved if the derived \re\ was larger than the HWHM of the corresponding PSF. In total we have 374 galaxies with resolved \ha\ emission. For the rest 63 unresolved sources, we adopt a size upper limit equal to the HWHM of the PSF.

Finally, we complement  our \ha\ size measurements with stellar size estimates by  matching our sample with the morphological catalog of Van 
der Wel et al.\,(2012). This catalog, produced using GALFIT (Peng et al.\,2002), offers, among other parameters, half light radius and sersic 
index measurements based on high resolution (0.17") HST-WFC3 $H_{\rm F160W}$ filter observations, that trace the stellar continuum of the 
galaxies at $z\sim 0.8-1.0$ and therefore can serve as stellar size estimates. Out of the whole sample, we have both 
stellar and \ha\ size measurements for a sub-sample of 117 galaxies in the COSMOS, GOODS-S and UDS fields. For completeness, and to 
investigate possible biases introduced by the derivation of the stellar sized based on the $H_{\rm F160W}-$band observations, we also 
consider the GALFIT morphological analysis of H{\"au}{\ss}ler et al.\,(2007) in ECDFS and Tasca et al.\,(2009) in COSMOS, based on (0.1") 
HST-ACS F814W imaging. These studies offer structural information for 182 galaxies in our sample.
\begin{figure*}
\centering
\includegraphics[scale=0.29]{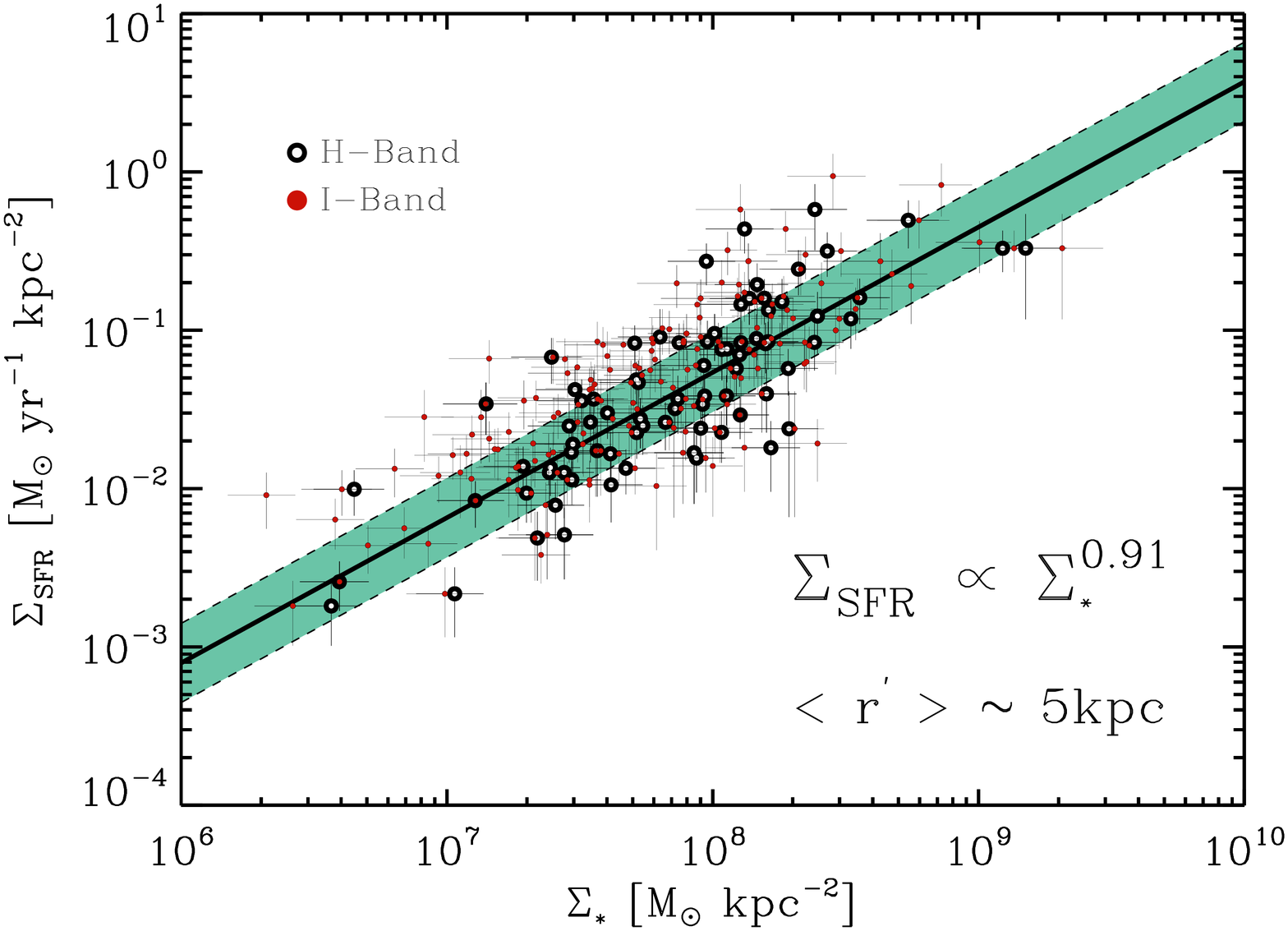}
\includegraphics[scale=0.29]{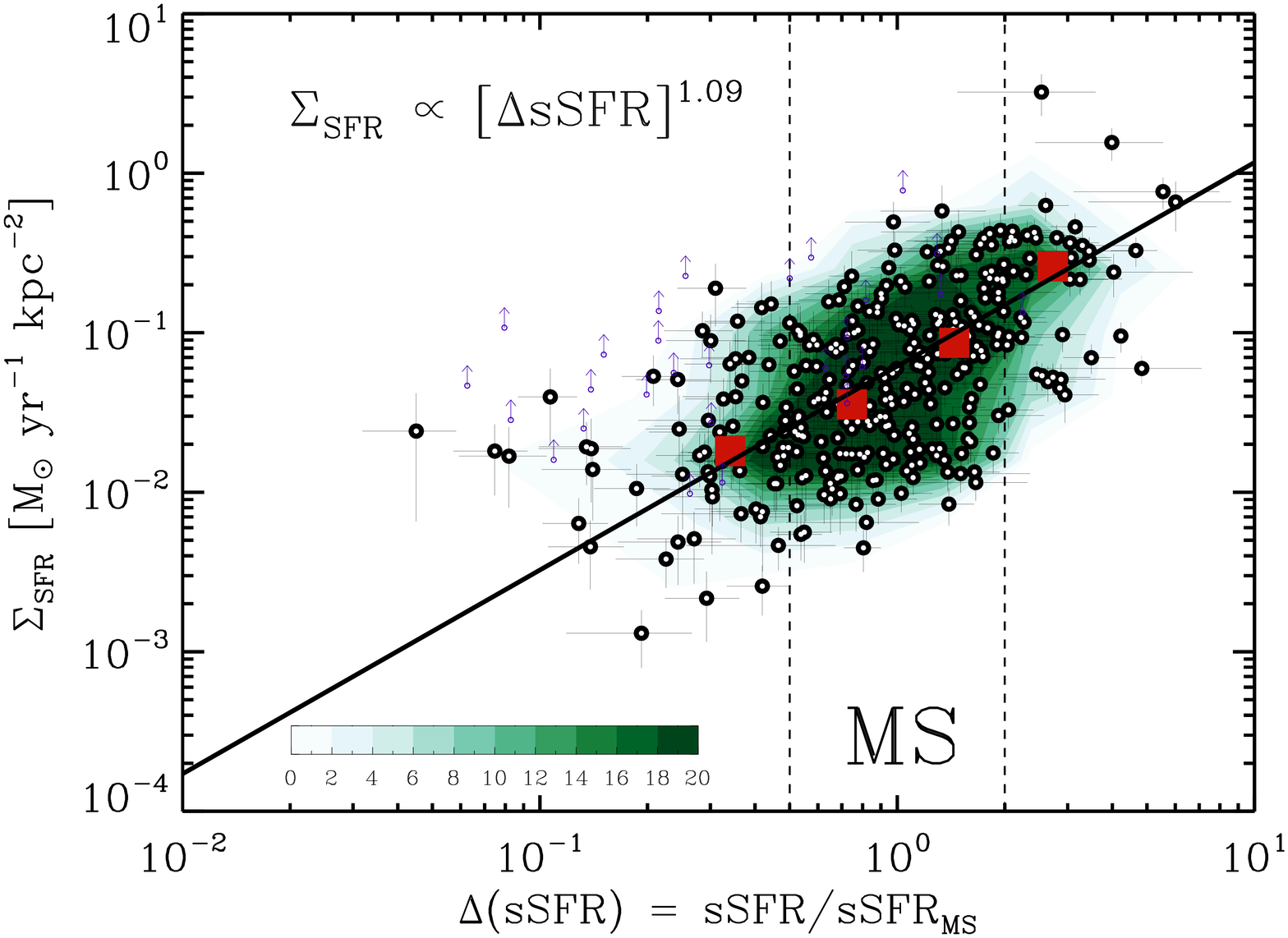}
\caption{\textbf{Left)} A ``resolved" (down to, on average, $\approx$ 5.0\,kpc) version of the star formation main sequence, showing the projected star formation rate density as a function of stellar density. It appears that the correlation between the ongoing and past star formation activity, as observed in the integrated properties of the galaxies holds at sub-galactic scales. The black solid and dashed lines represent the best fit to the data and the corresponding scatter (0.25\,dex). \textbf{Right)} Projected star formation rate density as a function of distance from the main sequence. The black line corresponds  best fit to the data, with a slope 1.09. The red boxes corresponds to the mean \den\ in four $\rm \Delta(sSFR)$ bins ($0.1 \leq \Delta(sSFR) < 0.5$, $0.5 \leq \Delta(sSFR) <1$, $1 \leq \Delta(sSFR) \leq 2$ and $2 < \Delta(sSFR) < 10$). The vertical dashed lines enclose the area of the main sequence, while the green contours depict the number density of the sources in the plot. Purple arrows correspond to lower limits for spatially unresolved sources with \ha\ detection. This plot demonstrates that, on average, galaxies with elevated sSFR with respect to the MS, are characterised by denser star formation activity.}
\label{fig:den1} %
\end{figure*}
\subsection{Metallicity}
Using the [\ion{N}{ii}] to \ha\ emission line ratio we are also in position to estimate the gas phase abundance of Oxygen [12+log(O/H)] (e,g, 
Kewley \& Dopita\,2002) in $\sim$ 190 galaxies from our sample for the  which we have both [\ion{N}{ii}] and  \ha\ detections. Using the Pettini 
\& Pagel\,(2004) formula: 12+log[O/H] = 8.9+0.57 $\times$ N2, where N2$ = log(f_{[NII]}/f_{\rm H\alpha})$, we derive individual metallicity 
estimates for each galaxy as well as an average metallicity for the sample of 8.64$\pm$ 0.02, that is very close to solar metalliciy. 

The lower detection rate of [\ion{N}{ii}] 
($\approx$ 40\%) compared to that of \ha\ ($\approx$ 90\%), could imply a bias of the sample towards lower metallicity sources. However, the non-detection of 
[\ion{N}{ii}], could be affected by imperfect subtraction of strong OH sky lines (e.g., Stott et al. 2013). Indeed, by measuring the distance of the [\ion{N}{ii}] central wavelength to that of strong ($>$20 relative strength) sky lines (Rousselot et al.\,2000), we find an average distance of $\approx$7.5\AA\ for the galaxies that have no detection in  [\ion{N}{ii}]. On the other hand, the average distance of the [\ion{N}{ii}] central wavelength for galaxies with [\ion{N}{ii}] detection is 11.7\,\AA. Given that the spectral resolution of KMOS in these wavelengths is $\sim$ 3.5\,\AA\, it is likely that a large fraction of non-[\ion{N}{ii}] detections are affected from the presence of skylines. We find that among the 247 galaxies with no [\ion{N}{ii}] detection, for 113 the [\ion{N}{ii}] wavelength is less than 7\,\AA\ (2$\times$FWHM) away from a strong skyline. These sources are subsequently omitted  from the metallicity analysis. For the  rest 134 galaxies with no [\ion{N}{ii}] but for which the [\ion{N}{ii}] wavelength is more than 7\,\AA\ (2$\times$FWHM) away from a strong skyline, we derive a metallicity upper limit assuming a 5$\sigma$ [\ion{N}{ii}] flux density, as derived from the noise of the spectrum at the  corresponding wavelength.

\section{Results and Discussion}
The main goal of this paper is to investigate if and how the width of the MS, as depicted in Figure \ref{fig:sfr}\,(right), as well as the deviation from the MS, reflect variations in the physical properties of the galaxies. With key parameters at hand, such as SFR, stellar, physical size, metallicity and stellar mass, for a large sample of $z\sim 1$ star forming galaxies, in this section we will first look into a resolved flavour of the MS and then proceed to explore possible variations of the star-formation surface density and the metal abundance of the galaxies as we move along and above the main sequence of star formation.

\subsection{A Resolved Main Sequence of Star Formation}

By combining the \ha\ based star-formation rates with the \ha\ sizes, we can derive the star formation surface density (\den), i.e. the amount of star formation rate per unit area, of our sources. We define the star formation surface density as: 

\begin{equation}
\begin{centering}
{  {\rm \Sigma_{SFR} = SFR}/2\pi r^{2}_{1/2}} \rm ~~~[M_{\odot}\,yr^{-1}\,Mpc^{-2}]
\end{centering}
\end{equation} 

Similarly, for the sub-sample of our sources with available high resolution HST-WFC3 $H_{\rm F160W}-$band imaging, we infer their stellar mass density (\mden), using the morphological analysis of Van der Wel et al.\,(2012). In particular, using the Sercic index, the half-light radius ($r_{\rm H}$) and the $H_{\rm F160W}-$band magnitude of the sources, as derived by Van der Wel et al.\,(2012) we infer the amount of light, and subsequently the stellar mass within the \ha\ half light radius of the galaxies, by assuming a constant mass to light ratio and rescaling the total stellar mass as derived from the galaxy integrated SED-fitting. An obvious caveat is that while the $H_{\rm F160W}-$band (rest frame 0.8\,$\mu$m at $z \sim 1$)  is in general sensitive to the stellar emission, the distribution of the stellar mass could be different from that of the $H_{\rm F160W}-$band light, especially in high resolution images. For example Wuyts et al.\,(2013), find that at a 0.7\,kpc resolution, the stellar mass of $z \sim1.0$ star forming galaxies  is more centrally concentrated than the $H_{\rm F160W}-$band light profile.  However, at a scale of  5.0\,kpc (which is the median \rei\ of the galaxies in our sample)  we expect this effect to be less pronounced. In fact, we find that  the \ha\ half light radius is, on average, 1.3 times larger than that of the  stellar light ($r_{\rm H}$) (in agreement with Nelson et al.\,2012). This means that we rescale the total stellar mass to that embedded within a radius $R > r_{\rm H}$. Therefore, even if the stellar distribution is more centrally concentrated, we integrate over a significant portion of the galaxy, smoothing out, to a large degree at least, possible variations in the M/L.

In Figure \ref{fig:den1}(left) we plot \den\ versus \mden\ showing a ``resolved'' version, of the well established galaxy-integrated  SFR$-$M$_{\ast}$ relation. Clearly, the correlation between the ongoing and past star formation activity appears to hold in sub-galactic scales, down to $\sim$ 5.1\,kpc (which is the average \ha\ half light radius of our sample), with a slightly sub-linear slope: 

\begin{equation}
\begin{centering}
{  {\rm log(\Sigma_{SFR} [M_{\odot}\,yr^{-1}\,kpc^{-2}]) = -8.21 + 0.91log(\Sigma_{*}[M_{\odot}\,kpc^{-2}] }}
\end{centering}
\end{equation} 

To test against any possible bias introduced by the $H_{\rm F160W}-$band observations. we also repeat the analysis using this time stellar 
size estimates based on ACS $F840W-$band imaging. As shown in Figure \ref{fig:den1}(left), the trend between \den\  and \mden\ 
remains overall unaffected by the adopted band used to infer stellar size measurements. 

The  correlation between (\den) and (\mden) that we derive here, is in excellent agreement with the slope of the galaxy-integrated main 
sequence (e.g. Whitaker et al.\,2012), albeit without the  flattening at the high mass end, possibly due to the absence of galaxies with high 
stellar density in our sample. A flattening in the high stellar density end of the  resolved MS is however reported in  Wuyts et al.\,(2013), that 
used 3D-HST and CANDELS data for a sample of $z \sim$ 1 star forming galaxies. While the slope of the resolved MS reported by Wuyts et al.\,(2013) is 
very similar to that obtained here,  we stress that  the resolution of our data is $\sim 5$ coarser than that of Wuyts et al.\,(2013) and any 
comparison between the two studies should be treated with caution. We conclude that the past and ongoing star formation follow closely each 
other not only in galactic but also in sub-galactic scales, suggesting that over the same area, the rate of star-formation tracks the amount of 
assembled stellar mass.

\subsection{Star-formation rate surface density along the MS}
The \den\ in our sample spans approximately 2 orders of magnitude, suggesting a wide range of combinations between SFR and \rei\ for our galaxies.
Several studies have attempted to investigate these variations of \den\ as a function of specific star formation rates or, equally of distance to the MS. For example Wuyts et al.\ (2011), using stellar continuum sizes report a correlation between \den\ and sSFR. Similarly, Elbaz et al.\ (2011), using radio size measurements, find a positive correlation between the two parameters, in the sense that galaxies with higher \den\ tend to be have higher IR8 and sSFR. Here, we are in position to address this question using the \ha\ size measurements, that directly probe the area in which star formation is taking place, of $\sim$400 galaxies with resolved \ha\ emission.

In Figure \ref{fig:den1}(right) we plot the derived star formation surface densities as a function of the distance to the MS ($\rm \Delta(sSFR)$), by comparing the sSFRs of the sources to a characteristic sSFR$_{\rm MS}$ that varies with stellar mass and redshift of each source,  following the Speagle et al.\ (2014) formula. This plot brings together the global star formation per unit stellar mass (sSFR) with the density of the star activity (\den) and therefore can provide information about the geometry and the mode of star formation (stabursting). For unresolved sources with \ha\ measurement, we consider a lower limit of their \den\ by adopting an upper \rei, equal to the resolution element of the corresponding observation. We find a significant correlation (Spearman's test $p-$value = $3\times 10^{-19}$) between \den\ and $\rm \Delta(sSFR)$, with increasing (decreasing) star formation surface densities as we depart above (below) the MS. In particular a least square regression fit to the data yield: \den\ $\propto \rm \Delta(sSFR)^{1.09}$  suggesting, on average,  a variation in the star formation density of a factor of $\sim$ 4 within the width of the MS (0.3\,dex).   

The increase of \den\ for increasing SFR at a fixed stellar mass, suggests that galaxies form more stars per unit area as we move above the MS. This 
 can be attributed to either an increase in the star formation efficiency (SFE=SFR/M$_{\rm gas}$),  or an increase in the gas fraction of the galaxies $f_{\rm gas}$, or a combination of the two. Various studies have shown that at fixed stellar mass, the gas fraction of MS galaxies increases for increasing SFR (e.g. Magdis et al.\,2012, Tacconi et al.\,2013, Genzel et al.\,2015). For example Magdis et al.\ (2012) report a trend of 
increasing M$_{\rm gas}$/M$_{\ast}$ (or equally of $f_{\rm gas}$) with increasing $\rm \Delta(sSFR)$, with a slope of $\xi \approx 0.9$. On the other 
hand, Saintonge et al. (2011) and Genzel et al.\,(2015) find that as galaxies above the MS are characterised by shorter gas depletion time 
scales, suggesting that the width of the MS is a combination of higher gas fractions and higher SFEs. While in the current study we cannot 
disentangle between the two processes, as we lack M$_{\rm gas}$ measurements, we confirm that measuring the sSFR of the galaxies with respect to the $\rm sSFR_{MS}$ carries important information regarding the star formation activity in the galaxies and that the width of the MS is not just an artefact produced by random noise, but a manifestation of the variation of the physical properties of the MS galaxies.

 Finally, we also find that starbursting galaxies are on average more compact than the MS galaxies, with  a mean size of 4.2 $\pm$0.5\,kpc,  and  5.2 $\pm$0.3\,kpc respectively. This trend, i.e.,  starbursting being more compact than MS galaxies, is consistent with Elbaz et al.\,(2011) who used ACS $V-$band stacks (i.e. rest frame UV for $z\sim1$ galaxies). As discussed in Section 2, unresolved galaxies in our sample are predominantly passive galaxies (red $r-z$ colours, high stellar masses, low SFRs) which typically tend to have more concentrated light profiles. Therefore, it is not surprising to find a lot of unresolved galaxies lying below the MS. On the other hand, since starburst galaxies are on average smaller than MS galaxies, one could expect to see some unresolved systems above the MS too. However, Elbaz et al.\,(2011) report an average UV size for starbursting systems ($ \rm \Delta(sSFR) >2 $) of 3.3\,kpc, dropping down to 2.5\,kpc when only strong starbusts are considered ($ \rm \Delta(sSFR) >3$). In both cases, the average size of galaxies above the MS is well above the typical resolution achieved in KROSS. This, along with the fact that our sample lacks strong outliers above the MS (almost all galaxies have $ \rm \Delta(sSFR) < 4$), could explain why all starbursts in our sample are resolved.
 
\subsection{Metallicity gradient along the MS}
Finally, using the gas phase abundance of Oxygen [12+log(O/H)] in our galaxies, as derived in section 3, we can investigate possible trends between the metallicity and the distance from the MS. Indeed, Ellison et al.\,(2008), were the first to show an anti-correlation between metallicity and sSFR at fixed stellar mass. Follow up studies, using SDSS data have further examined this trend, suggesting that the scatter around the well established metallicity $-$ stellar mass relation is correlated to the SFR of the galaxies (e.g. Yates et al.\,2012; Andrews \& Martini 2013, Salim et al.\,2014). To this extend, Mannucci et al.\,(2010), established a fundamental, redshift-invariant, relation between SFR-M$_{\ast}$-Z, the so called FMR relation, showing that at fixed stellar mass the metallicity of the galaxies decreases for increasing SFR. More recently, Salim et al.\,(2014), also presented evidence that metallicity is anti-correlated with sSFR regardless of the metallicity indicators used. Here, we can extend this investigation to high$-z$ galaxies (see also Stott et al.\, 2013).
\begin{figure}
\centering
\includegraphics[scale=0.28]{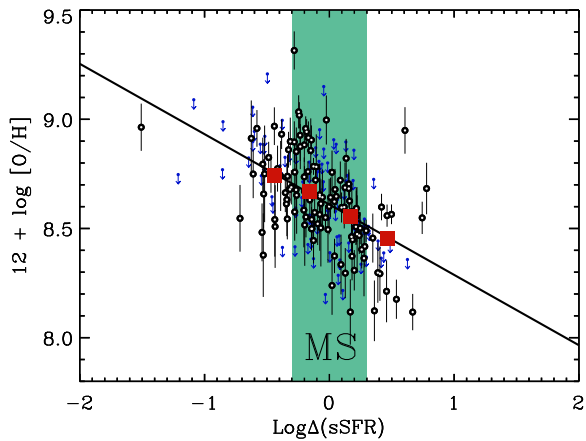}\\
\caption{Metallicity versus distance from the main sequence. The red boxes correspond to the mean metallicity values in four $\rm \Delta(sSFR)$ bins; one below, two within and one above the MS (same as in Figure 3right). The green shaded region depicts the locus of main sequence. The solid line represents the best fit to the data, with a slope of -0.36, suggesting a variation of 0.2\,dex in the metallicity of the galaxies within the MS and significantly lower metallicities for star bursting galaxies that sit well above the MS.}
\label{fig:co_spec} %
\end{figure}

In Figure 4, we plot the metallicity  as a function of distance from the MS, for galaxies in the stellar mass range of 9.5 $<$ log($ \rm M_{\ast}/
M_{\odot}$) $<$ 10.5. The stellar mass bin was adopted in order to minimise the dependence on the stellar mass. Evidently, and in agreement with 
what is observed in the local Universe, at fixed stellar mass, galaxies with higher SFR (or for equally galaxies departing from the MS) are found to have lower metallicities. A linear regression fit to the data yields: Z $\propto$ $(\rm \Delta(sSFR))^{-0.32}$, suggesting a variation of $\sim$0.2\,dex within the MS, and significantly lower metallicities for starbursting galaxies that sit well above the MS.

The observed trend can be explained via gas flows: galaxies with higher SFR at a fixed stellar mass are likely to experience higher accretion rates of 
pristine gas that dilute the metallicity content of the galaxies. For example in the model of Finlator \& Dave\,(2008), it is proposed that for a 
given stellar mass, all galaxies have an equilibrium metallicity, and deviations from that are caused by inflow of pristine gas from the IGM. In this 
scenario, we would expect higher gas surface densities and therefore higher \den, something that is validated in the previous analysis. To this direction,  Stott et al.\, (2014)  find a correlation between the gaseous metallicity gradient and sSFR of a galaxy using the KMOS SV data and literature measurements. They find that central metallicity is lower in high sSFR galaxies suggesting that this pristine gas is funnelled towards the core of these galaxies enhancing their SFRs. Furthemore, Bothwell et al.\,(2013) and De Rossi et al.\,(2015) presented evidence that the mass$-$metallicity relation exhibits a strong secondary dependence on total gas mass, rather than on SFR, with galaxies with higher gas fraction having lower metallicities. In this case, the deviation from the MS reflects variations in the gas fraction of the galaxies, as already discussed in the previous sub-section. On the other hand, as argued by Ellison et al.\,(2008), the observed trend could also be driven by star formation efficiency variations; higher gas densities could 
have lead to higher star formation efficiencies in the past ($z > 1$) that would yield higher metallicities for galaxies with lower SFR by $z = 1$. Again, as for the \den\ $-$ $\rm \Delta(sSFR)$ relation, while we recover a clear correlation between Z and $\rm \Delta(sSFR)$, we need $\rm M_{\rm gas}$ measurements in order to understand the physical mechanisms behind  the observed trends. 

Finally, Ellison et al.\,(2008) and Salim et al.\,(2008), provide evidence that at given stellar mass larger galaxies have on average smaller metallicities, suggesting a secondary  (but weaker to sSFR) dependence of the gas metallicity on the physical size of the galaxies. However, when we split  our sub-sample of resolved galaxies with a gas metallicity measurement in various stellar mass bins, we do not recover a statistically  significant correlation (or anti-correlation thereof)  between the gas metallicity and \rei. However, we note that the lack of anti-correlation between these two parameters is probably driven by 1) the small number of galaxies that populate in each stellar mass bin and 2) the fact that the range of the physical sizes of the galaxies in our sample is much narrower than that of the aforementioned studies (80\% of the galaxies with measured metallicities have 3\,kpc $<$ \rei\ $<$ 7\,kpc). Full treatment of the metallicity of the KROSS sample will be presented in Stott et al.\, (2016, in prep).
 
\section{Summary}
We have presented first results from the ongoing KROSS survey, that targets $\sim$1000 star forming galaxies at $z \sim 1$ with KMOS in order to detected and resolve their \ha\ emission. Using \ha\ flux densities and \ha\ size measurements as well as [\ion{N}{ii}] integrated flux densities for $\sim$500 observed galaxies so far, we have investigated various physical properties with respect to the main sequence of star formation. The main findings are as follows: 

\begin{itemize}

\item Based on \ha\ flux density measurements we derive SFR estimates for galaxies in our sample and recover the SFR-$\rm M_{\ast}$ relation at $z\sim 1$ with a scatter of 0.3\,dex, in agreement with previous studies.
\item Using \ha\ maps we measure the half light radius of the galaxies and present a ``resolved" version of the main sequence of star formation that appears to hold with the same slope and scatter in sub-galactic scales.
\item We find a strong correlation between the \den\ and the distance from the main sequence, parametrised as $\rm \Delta(sSFR)$ = $ \rm sSFR/sSFR_{\rm MS}$; galaxies tend to be characterised by denser star formation as we move above the MS. 
\item We find a clear trend between the metallicity of the galaxies and their SFR at a fixed stellar mass (or equally their distance form the MS). The recovered relation suggests lower metallicities for galaxies well above the MS and  a metallicity gradient of $\sim$0.2\,dex within the MS.
\end{itemize}

Follow up observations of the current sample to derive $\rm M_{gas}$ estimates, either through sub-mm continuum observations or direct CO detections will enable us to understand the physical origin of the above trends.

\section{Acknowledgements}
We would like to thank the referee for  carefully   reading   the manuscript  and  providing  valuable  comments  and  suggestions. 
G.E.M. acknowledges support from STFC through grant ST/K00106X/1, the John Fell Oxford University Press (OUP) Research Fund and the University of Oxford. JPS, CMH and IRS acknowledge support from STFC (ST/I001573/1). JPS also acknowledges support from a Hintze Research Fellowship. IRS acknowledges support from an ERC Advanced Investigator programme DUSTYGAL and a Royal Society/Wolfson Merit Award. AJB gratefully acknowledges the hospitality of the Research School of Astronomy \& Astrophysics at the Australian National University, Mount Stromlo, Canberra where some of this work was done under the Distinguished Visitor scheme. DS acknowledges financial support from the Netherlands Organisation for Scientific research (NWO) through a Veni fellowship and from FCT through a FCT Investigator Starting Grant and Start-up Grant (IF/01154/2012/CP0189/CT0010). We thank Ian Smail for co-ordinating the efforts of the KROSS survey and for stimulating discusssions. We thank Holly Elbert and Timothy Green for their observations and Matthieu Schaller for providing the EAGLE simulation quantities. GEM thanks Ryan Houghton and Borris Haussler for helpfull  discussions. Based on observations made with ESO Telescopes at the La Silla Paranal Observatory under the programme IDs 60.A-9460, 092.B-0538, 093.B-0106 and 094.B-0061.This research uses data from the VIMOS VLT Deep Survey, obtained from the VVDS database operated by Cesam, Laboratoire dÕAstrophysique de Marseille, France.This paper uses data from the VIMOS Public Extragalactic Redshift Survey (VIPERS). VIPERS has been performed using the ESO Very Large Telescope, under the ``Large Programme" 182.A-0886. The participating institutions and funding agencies are listed at http://vipers.inaf.it. This paper uses data from zCOSMOS which is based on observations made with ESO Telescopes at the La Silla or Paranal Observatories under programme ID 175.A-0839. We thank D. Sobral and M. Cirasuolo for helping with the survey.


\begin{thebibliography}{}
\bibitem[Andrews \& Martini(2013)]{2013ApJ...765..140A} Andrews, B.~H., \& Martini, P.\ 2013, ApJ, 765, 140
\bibitem[Bothwell et al.(2013)]{2013MNRAS.433.1425B} Bothwell, M.~S., Maiolino, R., Kennicutt, R., et al.\ 2013, MNRAS, 433, 1425 
\bibitem[Brinchmann et al.(2004)]{2004MNRAS.351.1151B} Brinchmann, J., Charlot, S., White, S.~D.~M., et al.\ 2004, MNRAS, 351, 1151 
\bibitem[Bruzual \& Charlot(2003)]{2003MNRAS.344.1000B} Bruzual, G., \& Charlot, S.\ 2003, MNRAS, 344, 1000
\bibitem[Cardamone et al.(2010)]{2010ApJS..189..270C} Cardamone, C.~N., van Dokkum, P.~G., Urry, C.~M., et al.\ 2010, ApJS, 189, 270 
\bibitem[Chabrier(2003)]{2003ApJ...586L.133C} Chabrier, G.\ 2003, ApJL, 586, L133 
\bibitem[Cirasuolo et al.(2007)]{2007MNRAS.380..585C} Cirasuolo, M., McLure, R.~J., Dunlop, J.~S., et al.\ 2007, MNRAS, 380, 585
\bibitem[Daddi et al.(2007)]{2007ApJ...670..156D} Daddi, E., Dickinson, M., Morrison, G., et al.\ 2007, ApJ, 670, 156
\bibitem[Daddi et al.(2010)]{2010ApJ...713..686D} Daddi, E., Bournaud, F., Walter, F., et al.\ 2010, ApJ, 713, 686 
\bibitem[Daddi et al.(2010)]{2010ApJ...714L.118D} Daddi, E., Elbaz, D., Walter, F., et al.\ 2010, ApJL, 714, L118
\bibitem[Davies(2007)]{2007MNRAS.375.1099D} Davies, R.~I.\ 2007, MNRAS, 375, 1099 
\bibitem[De Looze et al.(2014)]{2014A&A...568A..62D} De Looze, I., Cormier, D., Lebouteiller, V., et al.\ 2014, A\&A, 568, A62 
\bibitem[De Rossi et al.(2015)]{2015MNRAS.452..486D} De Rossi, M.~E., Theuns, T., Font, A.~S., \& McCarthy, I.~G.\ 2015, MNRAS, 452, 486 
\bibitem[D{\'{\i}}az-Santos et al.(2010)]{2010ApJ...723..993D} D{\'{\i}}az-Santos, T., Charmandaris, V., Armus, L., et al.\ 2010, ApJ, 723, 993 
\bibitem[D{\'{\i}}az-Santos et al.(2013)]{2013ApJ...774...68D} D{\'{\i}}az-Santos, T., Armus, L., Charmandaris, V., et al.\ 2013, ApJ, 774, 68 
\bibitem[Elbaz et al.(2007)]{2007AA...468...33E} Elbaz, D., Daddi, E., Le Borgne, D., et al.\ 2007, A\&A, 468, 33 
\bibitem[Elbaz]{elbaz11}Elbaz, D., Dickinson, M., Hwang, H.S., et al. 2011, A\&A, 533, 119
\bibitem[Ellison et al.(2008)]{2008ApJ...672L.107E} Ellison, S.~L., Patton, D.~R., Simard, L., \& McConnachie, A.~W.\ 2008, ApJL, 672, L107 
\bibitem[Finlator \& Dav{\'e}(2008)]{2008MNRAS.385.2181F} Finlator, K., \& Dav{\'e}, R.\ 2008, MNRAS, 385, 2181
\bibitem[F{\"o}rster Schreiber et al.(2009)]{2009ApJ...706.1364F} F{\"o}rster Schreiber, N.~M., Genzel, R., Bouch{\'e}, N., et al.\ 2009, ApJ, 706, 1364
\bibitem[Garilli et al.(2008)]{2008A&A...486..683G} Garilli, B., Le F{\`e}vre, O., Guzzo, L., et al.\ 2008, A\&A, 486, 683 
\bibitem[Garilli et al.(2014)]{2014A&A...562A..23G} Garilli, B., Guzzo, L., Scodeggio, M., et al.\ 2014, A\&A, 562, A23 
\bibitem[Genzel et al.(2010)]{2010MNRAS.407.2091G} Genzel, R., Tacconi, L.~J., Gracia-Carpio, J., et al.\ 2010, MNRAS, 407, 2091 
\bibitem[Genzel et al.(2015)]{2015ApJ...800...20G} Genzel, R., Tacconi, L.~J., Lutz, D., et al.\ 2015, ApJ, 800, 20 
\bibitem[Grarcia-Carpio]{gc10}Gracia-Carpio, J., Sturm, E., Hailey-Dunsheath, S. et al. 2010, ApJ, 728, 7
\bibitem[Gonz{\'a}lez et al.(2010)]{2010ApJ...713..115G} Gonz{\'a}lez, V., Labb{\'e}, I., Bouwens, R.~J., et al.\ 2010, ApJ, 713, 115
\bibitem[Guzzo et al.(2014)]{2014A&A...566A.108G} Guzzo, L., Scodeggio, M., Garilli, B., et al.\ 2014, A\&A, 566, A108 
\bibitem[H{\"a}ussler et al.(2007)]{2007ApJS..172..615H} H{\"a}ussler, B., McIntosh, D.~H., Barden, M., et al.\ 2007, ApJS, 172, 615 
\bibitem[Hung et al.(2013)]{2013ApJ...778..129H} Hung, C.-L., Sanders, D.~B., Casey, C.~M., et al.\ 2013, ApJ, 778, 129
\bibitem[Kennicutt(1998)]{1998ARA&A..36..189K} Kennicutt, R.~C., Jr.\ 1998, ARA\&A, 36, 189 
\bibitem[Kewley et al.(2001)]{2001ApJS..132...37K} Kewley, L.~J., Heisler, C.~A., Dopita, M.~A., \& Lumsden, S.\ 2001, ApJS, 132, 37 
\bibitem[Kewley \& Dopita(2002)]{2002ApJS..142...35K} Kewley, L.~J., \& Dopita, M.~A.\ 2002, ApJS, 142, 35
\bibitem[Kim(2002)]{2002ApJS..142...35K}Kim J.-W., Edge A. C., Wake D. A., Stott J. P., 2011b, MNRAS,410, 241 
\bibitem[Larson et al.(2011)]{2011ApJS..192...16L} Larson, D., Dunkley, J.,Hinshaw, G., et al.\ 2011, ApJS, 192, 16
\bibitem[Lawrence et al.(2007)]{2007MNRAS.379.1599L} Lawrence, A., Warren, S.~J., Almaini, O., et al.\ 2007, MNRAS, 379, 1599
\bibitem[Le F{\`e}vre et al.(2005)]{2005A&A...439..845L} Le F{\`e}vre, O., Vettolani, G., Garilli, B., et al.\ 2005, A\&A, 439, 845 
\bibitem[Le F{\`e}vre et al.(2013)]{2013A&A...559A..14L} Le F{\`e}vre, O., Cassata, P., Cucciati, O., et al.\ 2013, A\&A, 559, A14 
\bibitem[Lehmer et al.(2005)]{2005ApJS..161...21L} Lehmer, B.~D., Brandt, W.~N., Alexander, D.~M., et al.\ 2005, ApJS, 161, 21
\bibitem[Lilly et al.(2007)]{2007ApJS..172...70L} Lilly, S.~J., Le F{\`e}vre, O., Renzini, A., et al.\ 2007, ApJS, 172, 70   
\bibitem[Magdis et al.(2010)]{2010MNRAS.401.1521M} Magdis, G.~E., Rigopoulou, D., Huang, J.-S., \& Fazio, G.~G.\ 2010, MNRAS, 401, 1521 
\bibitem[Magdis et al.(2012)]{2012ApJ...760....6M} Magdis, G.~E., Daddi, E., B{\'e}thermin, M., et al.\ 2012, ApJ, 760, 6
\bibitem[Magdis et al.(2014)]{2014ApJ...796...63M} Magdis, G.~E., Rigopoulou, D., Hopwood, R., et al.\ 2014, ApJ, 796, 63
\bibitem[Magnelli et al.(2014)]{2014A&A...561A..86M} Magnelli, B., Lutz, D., Saintonge, A., et al.\ 2014, A\&A, 561, AA86
\bibitem[Mancini et al.(2011)]{2011ApJ...743...86M} Mancini, C., F{\"o}rster Schreiber, N.~M., Renzini, A., et al.\ 2011, ApJ, 743, 86 
\bibitem[Mannucci et al.(2010)]{2010MNRAS.408.2115M} Mannucci, F., Cresci, G., Maiolino, R., Marconi, A., \& Gnerucci, A.\ 2010, MNRAS, 408, 2115
\bibitem[Muzzin et al.(2013)]{2013ApJS..206....8M} Muzzin, A., Marchesini, D., Stefanon, M., et al.\ 2013, ApJS, 206, 8 
\bibitem[Nelson et al.(2012)]{2012ApJ...747L..28N} Nelson, E.~J., van Dokkum, P.~G., Brammer, G., et al.\ 2012, ApJL, 747, L28
\bibitem[Noeske et al.(2007)]{2007ApJ...660L..43N} Noeske, K.~G., Weiner, B.~J., Faber, S.~M., et al.\ 2007, ApJL, 660, L43
\bibitem[Nordon et al.(2012)]{2012ApJ...745..182N} Nordon, R., Lutz, D., Genzel, R., et al.\ 2012, ApJ, 745, 182
\bibitem[Pannella et al.(2009)]{2009ApJ...698L.116P} Pannella, M., Carilli, C.~L., Daddi, E., et al.\ 2009, ApJL, 698, L116 
\bibitem[Peng et al.(2002)]{2002AJ....124..266P} Peng, C.~Y., Ho, L.~C., Impey, C.~D., \& Rix, H.-W.\ 2002, AJ, 124, 266
\bibitem[Pettini \& Pagel(2004)]{2004MNRAS.348L..59P} Pettini, M., \& Pagel, B.~E.~J.\ 2004, MNRAS, 348, L59 
\bibitem[Saintonge et al.(2011)]{2011MNRAS.415...61S} Saintonge, A., Kauffmann, G., Wang, J., et al.\ 2011, MNRAS, 415, 61 
\bibitem[Salim et al.(2014)]{2014ApJ...797..126S} Salim, S., Lee, J.~C., Ly, C., et al.\ 2014, ApJ, 797, 126 
\bibitem[Scoville et al.(2007)]{2007ApJS..172....1S} Scoville, N., Aussel, H., Brusa, M., et al.\ 2007, ApJS, 172, 1 
\bibitem[Schreiber et al.(2015)]{2015A&A...575A..74S} Schreiber, C., Pannella, M., Elbaz, D., et al.\ 2015, A\&A, 575, A74 
\bibitem[Sharples et al.(2013)]{2013Msngr.151...21S} Sharples, R., Bender, R., Agudo Berbel, A., et al.\ 2013, The Messenger, 151, 21 
\bibitem[Sharples et al.(2013)]{2013Msngr.151...21S} Simpson J. M. et al., 2014, ApJ, 788, 125
\bibitem[Smail et al.(2008)]{2008MNRAS.389..407S} Smail, I., Sharp, R., Swinbank, A.~M., et al.\ 2008, MNRAS, 389, 407 
\bibitem[Sobral et al.(2012)]{2012MNRAS.420.1926S} Sobral, D., Best, P.~N., Matsuda, Y., et al.\ 2012, MNRAS, 420, 1926 
\bibitem[Sobral et al.(2013)]{2013MNRAS.428.1128S} Sobral, D., Smail, I., Best, P.~N., et al.\ 2013, MNRAS, 428, 1128
\bibitem[Sobral et al.(2015)]{2015MNRAS.451.2303S} Sobral, D., Matthee, J., Best, P.~N., et al.\ 2015, MNRAS, 451, 2303 
\bibitem[Speagle et al.(2014)]{2014ApJS..214...15S} Speagle, J.~S., Steinhardt, C.~L., Capak, P.~L., \& Silverman, J.~D.\ 2014, ApJS, 214, 15
\bibitem[Steidel et al.(1998)]{1998ApJ...492..428S} Steidel, C.~C., Adelberger, K.~L., Dickinson, M., et al.\ 1998, ApJ, 492, 428 
\bibitem[Stott et al.(2013)]{2013MNRAS.430.1158S} Stott, J.~P., Sobral, D., Smail, I., et al.\ 2013, MNRAS, 430, 1158 
\bibitem[Stott et al.(2013)]{2013MNRAS.436.1130S} Stott, J.~P., Sobral, D., Bower, R., et al.\ 2013, MNRAS, 436, 1130 
\bibitem[Stott et al.(2014)]{2014MNRAS.443.2695S} Stott, J.~P., Sobral, D., Swinbank, A.~M., et al.\ 2014, MNRAS, 443, 2695
\bibitem[Tasca et al.(2009)]{2009A&A...503..379T} Tasca, L.~A.~M., Kneib, J.-P., Iovino, A., et al.\ 2009, A\&A, 503, 379 
\bibitem[Yates et al.(2012)]{2012MNRAS.422..215Y} Yates, R.~M., Kauffmann, G., \& Guo, Q.\ 2012, MNRAS, 422, 215 
\bibitem[van der Wel et al.(2012)]{2012ApJS..203...24V} van der Wel, A., Bell, E.~F., H{\"a}ussler, B., et al.\ 2012, ApJS, 203, 24 
\bibitem[Whitaker et al.(2012)]{2012ApJ...745..179W} Whitaker, K.~E., Kriek, M., van Dokkum, P.~G., et al.\ 2012, ApJ, 745, 179 
\bibitem[Whitaker et al.(2014)]{2014ApJ...795..104W} Whitaker, K.~E., Franx, M., Leja, J., et al.\ 2014, ApJ, 795, 104 
\bibitem[Wuyts et al.(2011)]{2011ApJ...738..106W} Wuyts, S., F{\"o}rster Schreiber, N.~M., Lutz, D., et al.\ 2011, ApJ, 738, 106 
\bibitem[Wuyts et al.(2013)]{2013ApJ...779..135W} Wuyts, S., F{\"o}rster Schreiber, N.~M., Nelson, E.~J., et al.\ 2013, ApJ, 779, 135
\end{thebibliography}
\end{document}